\documentclass{sig-alternate-05-2015}

\usepackage{epstopdf}
\usepackage{epsfig}
\usepackage{graphicx}
\usepackage{pdfpages}
\usepackage{caption}
\usepackage{subcaption}
\usepackage{listings}
\usepackage{url}
\usepackage{breakurl}

\usepackage{graphicx}
\usepackage{algorithmic}
\usepackage{listings}

\lstdefinelanguage{barriers}{%
  keywords={[1]for,do,while,if,else,endif,break,continue,return,
    typedef,struct,_Atomic},
  keywords={[2]const,volatile,static,signed,unsigned,__attribute__,%
    void,bool,char,short,long,int,float,double,boolean,size_t,sr_Bar_t, tp_Data_t},
  keywords={[4]thread_fence,load_explicit,store_explicit,%
    compare_exchange_strong_explicit,compare_exchange_weak_explicit,%
    relaxed,acquire,release,seq_cst,R,W,sync,ctrl_isync,atomic,mic_store,mic_pause,%
    add_and_fetch,_mm512_load_epi32,_mm512_div_epi32,_mm512_rem_epi32,_mm512_prefetch_i32gather_ps,
_mm512_i32gather_epi32,_mm512_sllv_epi32,_mm512_knot,_mm512_kor,_mm512_test_epi32_mask,
_mm512_mask_prefetch_i32scatter_ps,_mm512_mask_i32scatter_epi32,_mm512_mask_or_epi32, _mm512_set1_epi32},
  emph={main,sr_wait},
  string=[b]",
  comment=[l]//,
  morecomment=[s]{/*}{*/},
  flexiblecolumns=true,
  tabsize=2,
  captionpos=b,
  frame=single,
  framerule=0pt,
  aboveskip=1em,
  belowskip=1pt,
  framesep=0pt,
  basicstyle=\ttfamily\fontsize{7}{7}\selectfont,
  keywordstyle={[1]\bfseries},
  keywordstyle={[2]\ttfamily},
  keywordstyle={[3]\bfseries},
  keywordstyle={[4]\bfseries},
  emphstyle=\itshape,
  identifierstyle=\color{black},
  commentstyle=\color{darkgray}\itshape,
  stringstyle=\color{darkgray}
}

\usepackage{algorithm}

\begin{document}

% Copyright
\setcopyright{acmcopyright}
%\setcopyright{acmlicensed}
%\setcopyright{rightsretained}
%\setcopyright{usgov}
%\setcopyright{usgovmixed}
%\setcopyright{cagov}
%\setcopyright{cagovmixed}

% DO
\doi{10.475/123_4}

% ISBN
\isbn{123-4567-24-567/08/06}

%Conference
\conferenceinfo{PLDI '13}{June 16--19, 2013, Seattle, WA, USA}

\acmPrice{\$15.00}

%
% --- Author Metadata here ---
\conferenceinfo{WOODSTOCK}{'97 El Paso, Texas USA}
%\CopyrightYear{2007} % Allows default copyright year (20XX) to be over-ridden - IF NEED BE.
%\crdata{0-12345-67-8/90/01}  % Allows default copyright data (0-89791-88-6/97/05) to be over-ridden - IF NEED BE.
% --- End of Author Metadata ---

\title{Breadth First Search Vectorization on the Intel Xeon Phi}
%\subtitle{[Extended Abstract]
%\titlenote{A full version of this paper is available as
%\textit{Author's Guide to Preparing ACM SIG Proceedings Using
%\LaTeX$2_\epsilon$\ and BibTeX} at
\numberofauthors{3} %  in this sample file, there are a *total*
\author{
% 1st. author
\alignauthor Mireya Paredes\\
       \affaddr{University of Manchester}\\
       \affaddr{Oxford Road, M13 9PL}\\
       \affaddr{Manchester, UK}\\
       \email{{\normalsize paredesm@manchester.ac.uk}}
% 2nd. author
\alignauthor Graham Riley \\
       \affaddr{University of Manchester}\\
       \affaddr{Oxford Road, M13 9PL}\\
       \affaddr{Manchester, UK}\\
       \email{{\normalsize graham.riley@manchester.ac.uk}}
 %3rd. author
\alignauthor Mikel Luj\'an \\
       \affaddr{University of Manchester}\\
       \affaddr{Oxford Road, M13 9PL}\\
       \affaddr{Manchester, UK}\\
       \email{{\normalsize mikel.lujan@manchester.ac.uk}}
%\and  % use '\and' if you need 'another row' of author names
%% 4th. author
%\alignauthor Lawrence P. Leipuner\\
%       \affaddr{Brookhaven Laboratories}\\
%       \affaddr{Brookhaven National Lab}\\
%       \affaddr{P.O. Box 5000}\\
 %      \email{lleipuner@researchlabs.org}
%% 5th. author
%\alignauthor Sean Fogarty\\
%       \affaddr{NASA Ames Research Center}\\
%       \affaddr{Moffett Field}\\
%       \affaddr{California 94035}\\
%       \email{fogartys@amesres.org}
%% 6th. author
%\alignauthor Charles Palmer\\
%       \affaddr{Palmer Research Laboratories}\\
%       \affaddr{8600 Datapoint Drive}\\
%       \affaddr{San Antonio, Texas 78229}\\
%       \email{cpalmer@prl.com}
}
% There's nothing stopping you putting the seventh, eighth, etc.
% author on the opening page (as the 'third row') but we ask,
% for aesthetic reasons that you place these 'additional authors'
% in the \additional authors block, viz.
%\additionalauthors{Additional authors: John Smith (The Th{\o}rv{\"a}ld Group,
%email: {\texttt{jsmith@affiliation.org}}) and Julius P.~Kumquat
%(The Kumquat Consortium, email: {\texttt{jpkumquat@consortium.net}}).}
\date{30 July 1999}
% Just remember to make sure that the TOTAL number of authors
% is the number that will appear on the first page PLUS the
% number that will appear in the \additionalauthors section.
% There's nothing stopping you putting the seventh, eighth, etc.
% author on the opening page (as the 'third row') but we ask,
% for aesthetic reasons that you place these 'additional authors'
% in the \additional authors block, viz.

% Just remember to make sure that the TOTAL number of authors
% is the number that will appear on the first page PLUS the
% number that will appear in the \additionalauthors section.

\maketitle
\begin{abstract}
Breadth First Search (BFS) is a building block for graph algorithms
and has recently been used for large scale analysis of information in
a variety of applications including social networks, graph databases
and web searching. Due to its importance, a number of different parallel
programming models and architectures have been exploited to optimize
the BFS. However, due to the irregular memory access patterns and the
unstructured nature of the large graphs, its efficient parallelization is a 
challenge. The Xeon Phi is a massively parallel architecture
available as an \textit{off-the-shelf} accelerator, which includes a powerful
512 bit vector unit with optimized scatter and gather
functions. Given its potential benefits, work related to
graph traversing on this architecture is an active area of research.

We present a set of experiments in which we explore architectural
features of the Xeon Phi and how best to exploit them in a top-down
BFS algorithm but the techniques can be applied to the current
state-of-the-art hybrid, top-down plus bottom-up, algorithms.

We focus on the exploitation of the vector unit by developing an
improved highly vectorized OpenMP parallel algorithm, using vector
intrinsics, and understanding the use of data alignment and
prefetching. In addition, we investigate the impact of hyperthreading
and thread affinity on performance, a topic that appears under
researched in the literature. As a result, we achieve what we believe
is the fastest published top-down BFS algorithm on the version of Xeon
Phi used in our experiments. The vectorized BFS top-down source code
presented in this paper can be available on request as free-to-use software.

\end{abstract}




%
%  Use this command to print the description
%
%\printccsdesc

% We no longer use \terms command


\section{Introduction}

Today, scientific experiments in many domains, and large organizations
can generate huge amounts of data, e.g.\ social networks, health-care
records and bio-informatics applications \cite{graphCT}. Graphs seem to
be a good match for important and large dataset analysis as they can
abstract real world networks. To process large graphs, different
techniques have been applied, including parallel programming. The main
challenge in the parallelization of graph processing is that large
graphs are often unstructured and highly irregular and this limits
their scalability and performance when executing on
\textit{off-the-shelf} systems \cite{challenges}.

The Breadth First Search (BFS) is a building block of graph
algorithms. Despite its simplicity, its parallelization has been a
real challenge but it remains a good candidate for acceleration. To
date, different accelerators have been targeted to improve the
performance of this algorithm such as GPGPUs~\cite{Olukotun2011} and
FPGAs~\cite{Yaman15}. The Intel Xeon Phi, also known as MIC (Intel's
Many Integrated Core Architecture), is a massively parallel
\textit{off-the-shelf} system consisting of up to 60 cores with 4-way
simultaneous multi-threading (SMT) for a maximum of 240 logical cores
\cite{xeonphiBook2}. As well as this thread-level parallelism, each
core has a 512 bit vector unit allowing data-level parallelism to be
extracted by using Single Instruction Multiple Data (SIMD)
programming. We believe that by exploiting both of these forms of
parallelism, the Xeon Phi is an interesting platform for exploring
parallel implementations of graph algorithms.

The goal of this study is to demonstrate through experiments and
analysis the impact of using the Xeon Phi architecture to accelerate
BFS in a single Xeon Phi device, as a step towards the multi-device
solutions that will be needed to tackle very large graph-based
datasets. As a starting point, we took the description of the top-down
BFS algorithm in \cite{Gao2013} which is summarised in
Section~\ref{sec:BFS}. Although this offers a good starting point on
the Xeon Phi, there are still some architectural features that need to
be well understood in order to be exploited. The
contribution of this paper is twofold; first, we present
the results of a series of experiments demonstrating the benefit of
successfully exploiting architectural features of
the Xeon Phi focusing on the vector unit (programmed via vector
intrinsics), with data
alignment and prefetching. In addition, we present the results of
investigations into the impact of
hyperthreading (Intel's term for SMT) and thread affinity when the Phi
is underpopulated with threads.

The structure of the paper is as follows. The Xeon Phi architecture is
presented in Section \ref{sec:mic}. We present the procedure of
vectorizing the BFS algorithm on the Xeon Phi, starting with the
serial BFS algorithm in Section~\ref{sec:serBFS}, followed by an
initial parallel version in Section~\ref{sec:parBFS}, which we call
\textit{non-simd} version. Next, we discuss the \textit{simd} version
that exploits the vector unit on the Xeon Phi in
Section~\ref{sec:vectorization}. We then present performance
comparisons between the \textit{non-simd} and \textit{simd} versions
and explore the impact of parallelism at two levels of granularity
(OpenMP threading and the vector unit), achieving better results for
the native BFS top-down algorithm on the Xeon Phi than those
previously presented in \cite{MICgraphs} and \cite{stanic2014}. In
Section~\ref{sec:vectorization} we explore the impact of data
prefetching and discuss the effect of thread affinity mapping, leading
us to key optimizations, and motivating future work on the possibility
of using helper threads to improve performance. The experimental setup
and analysis of our results are shown in Sections~\ref{sec:exp}
and~\ref{sec:res}. Related work is briefly discussed in Section
\ref{relwork} and conclusions and future work are discussed in Section
\ref{sec:conc}.

\section{The Xeon Phi Architecture}
\label{sec:mic}
\begin{figure}
\centering
\includegraphics[scale=1.0, width=0.5\textwidth]{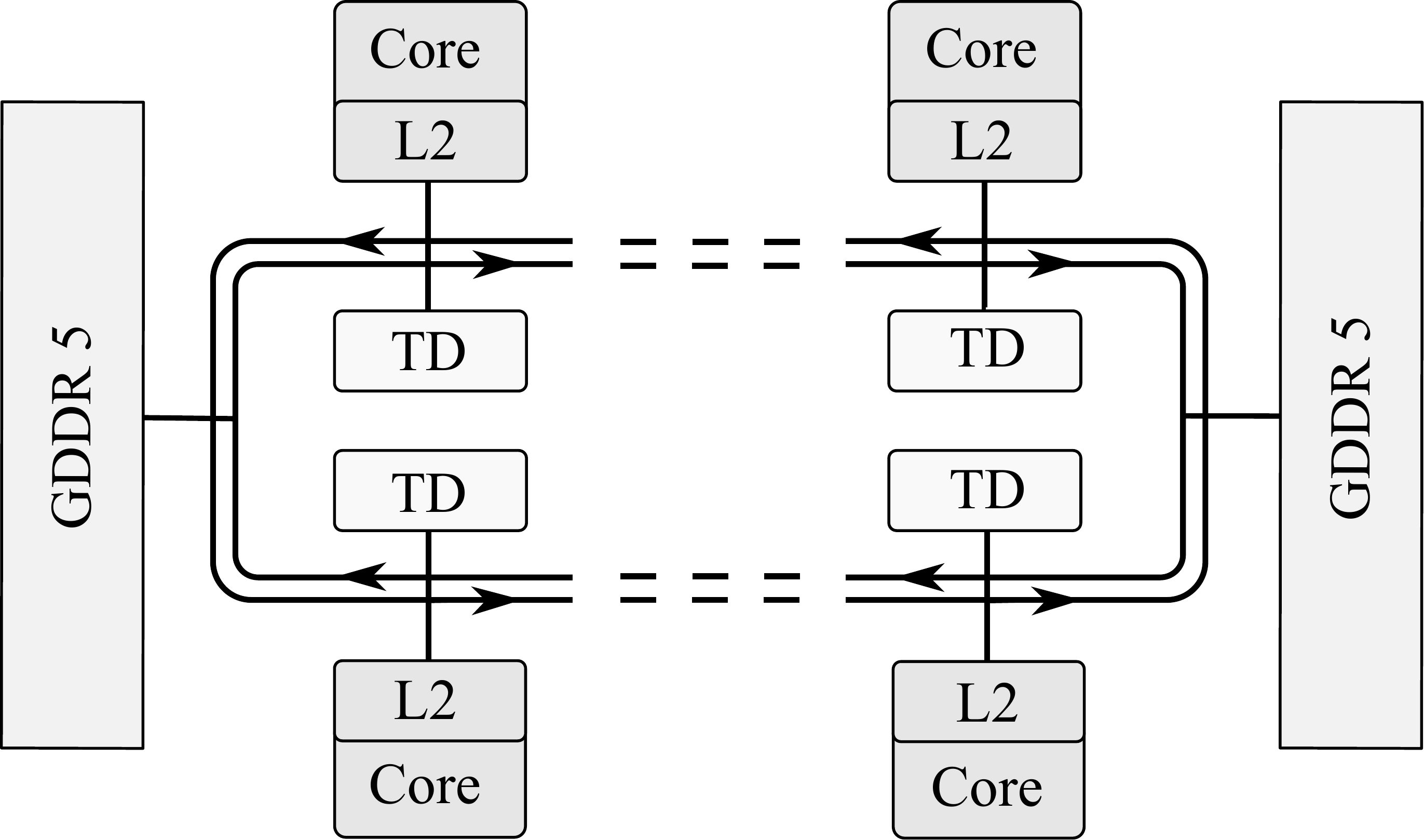}%mic_arch2.png}
\caption{The Intel \textsuperscript{\textregistered} Xeon Phi Microarchitecture.}
\label{fig:mic}
\vspace{-3mm}
\end{figure}

The Intel \textsuperscript{\textregistered} Xeon Phi
\textsuperscript{TM} coprocessor used in this work is composed of 60,
4 way-SMT Pentium-based cores \cite{xeonphiBook2} and a main memory of
8 GB. Each core contains a powerful 512-bits vector processing unit
and a cache memory divided into L1 (32KB) and L2 (512KB) kept fully
coherent by a global-distributed tag directory (TD), coordinated by
the cache coherency MESI protocol. Cores are interconnected through a
high-speed bi-directional ring bus \cite{xeonphiBook2} as it is shown
in Figure \ref{fig:mic}. Maximum memory bandwidth is quoted as 320GB/s.

% It allows to have a bandwidth of 320 GB per second. 

%single instruction multiple data (SIMD), that allows to process up to 16 elements of 32 bits at a time or 8 of 64 bits length and an indivual cache memory. 

The vector process unit (VPU) is composed of vector registers and
16-bit mask registers. Each vector register can process either 16
(32-bit) operations or 8 (64-bit) operations at a time. A vector mask
consists of 16 bits that control the update of the vector
elements. Only those elements whose bits are set to 1 are updated into
the vector register, the ones with 0 value in the mask remaining
unchanged. The Xeon Phi contains both hardware (HWP) and software
(SWP) prefetching, which in some cases can help to reduce memory
latency.

%HWP is activated by default, however if software prefetching is performing well then the compiler switchs to SWP.

The Xeon Phi can be programmed to support vectorization at two levels:
automatic and manual. In automatic vectorization the compiler
identifies and optimizes all parts of the code that can be vectorized
without the intervention of the programmer. However, there are some
obstacles that can limit the vector unit utilization, such as
non-contiguous memory accesses or data dependencies \cite{vecguide}
. In such cases, manual vectorization can be used allowing the user to
force the compiler to vectorize certain parts in the code. Manual
vectorization can be set by using SIMD pragma directives supported in
the compiler. The compiler also supports a wide range of {\em
  intrinsics} which allow a programmer low-level control of the vector
unit.

\section{Breadth First Search}
\label{sec:BFS}
 The Breadth First Search (BFS) algorithm is one of the building
 blocks for graph analysis algorithms including betweenness
 centrality, shortest path and connected components
 \cite{graphanalytics}. Given a graph $G$ and a starting vertex $s$,
 the BFS systematically explores all the edges $E$ of $G$ to trace all
 the vertices $V$ that can be reached from \textit{s} \cite{Cormen},
 where $|V|$ is the number of vertices and $|E|$ is the number of
 edges. The output is a BFS spanning tree ({\em bfs tree} of all the
 vertices encountered from $s$.

 %This format consists on having two arrays. Each index in the first array represents each row of the sparse matrix where each location contains the starting position of the column list, that every row is related with, in the second array. This matrix representation is commonly used for sparse graphs because it avoids non zero entries. Those are locations where there is not any relation between row and column, so there is no value. Thus, it saves storage but also time performance during its iteration. However, it might be only efficient to represent sparse graphs.

%The input graph data can be taken from different sources. Two of the sources that we are using are the University of Florida Sparse Matrix Collection \cite{UFS} and the Graph500 graph generator \cite{g500}. 

%The simplest sequential BFS algorithm consists of having a FIFO, which is a dynamic data structure with two operations: enqueue and dequeue. \textit{Enqueue} adds the next vertices to be processed at the end of the list. \textit{Dequeue} takes the first vertex out of the list to be processed. Both operations can be implemented in constant time $\Theta(1)$ \cite{leiserson}. Thus, the running time of this serial BFS algorithm is $O(V+E)$. 

The simplest sequential BFS algorithm consists in having a queue that
contains a list of vertices waiting to be processed. Enqueue and
dequeue operations on the queue can be implemented in constant time
$\Theta(1)$ \cite{leiserson}. Thus, the running time of this serial
BFS algorithm is $O(V+E)$. Despite the queue being simple and
efficient, during parallelization, it has some drawbacks. The queue
can be a bottleneck since it implies a vertex processing order because
the dequeue operation typically takes out the vertex that has been
added first. To address this problem vertices can be partitioned into
layers. A layer consists of a set of all vertices with the same
distance\footnote{Distance is the number of edges in the shortest path
  between two vertices.} from the source vertex. Processing vertices
by layers avoids the order restriction imposed by the queue, allowing
vertices to be explored in any order as long as they are in the same
layer. However, each layer has to be processed in sequence; that is
all vertices with distance $k$, (layer $L_{k}$) are processed before
those in layer $L_{k+1}$.

\begin{figure}
\centering
\includegraphics[width=0.5\textwidth, scale=0.2]{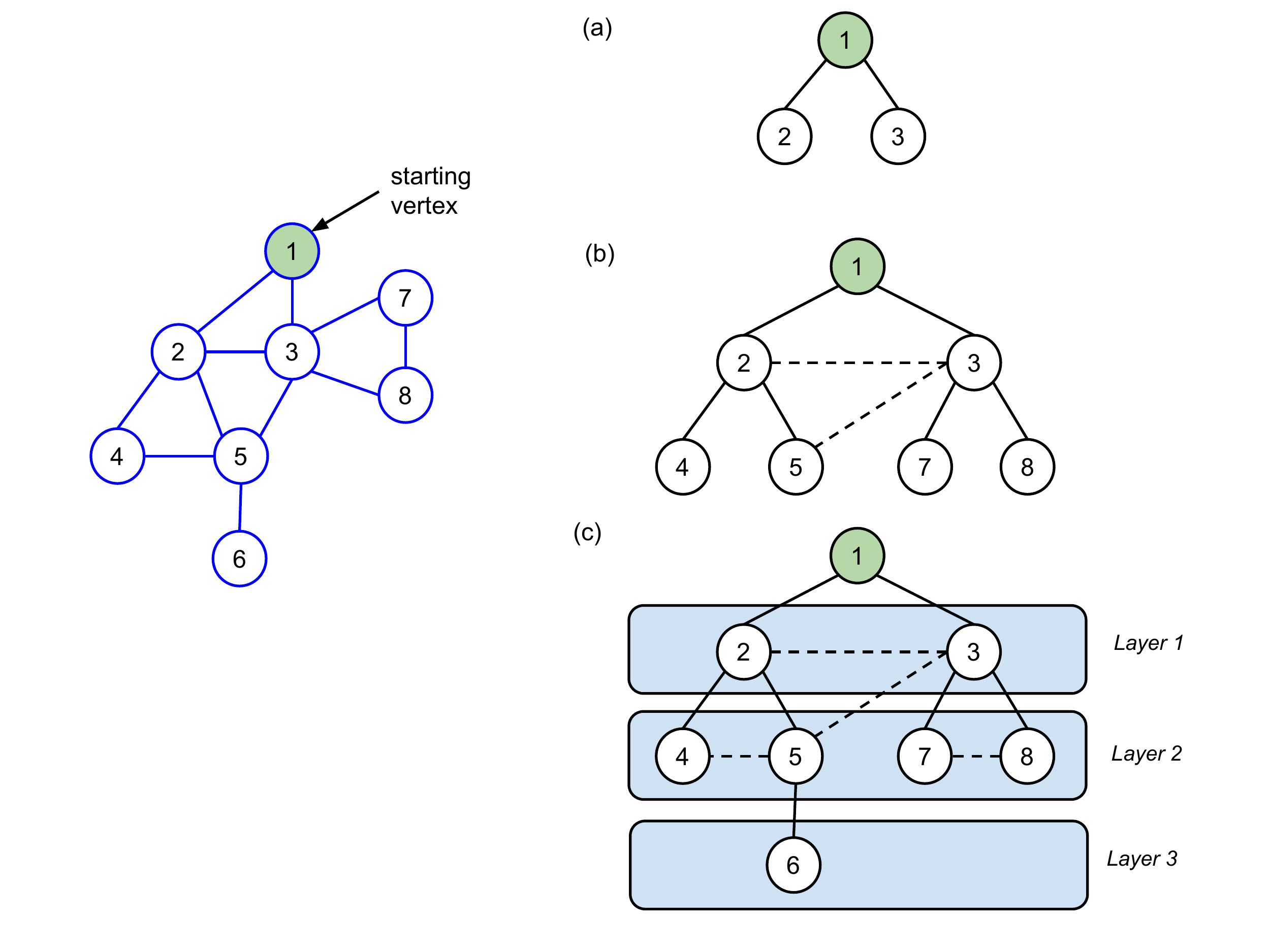}
\caption{An example of the Top-Down Breadth First spanning tree.}
\label{fig:bfslayers}
\vspace{-3mm}
\end{figure}

 Traversing the graph from the vertices in the top layer down to their
 children in the bottom layer is a conventional approach known as
 \textit{top-down} whereas traversing from the vertices in the bottom
 layer to find their parent in the top layer is known as
 \textit{bottom-up} approach. The \textit{bottom-up} concept was
 introduced by \cite{Beamer:2012} in a hybrid approach that explores
 the graph in both directions, first traversing the graph with the
 top-down and then swaping to the bottom-up approach in the middle layers
 to finally swap back to the top-down.
% Although the hybrid algorithm has shown better results than the top-down approach, we focus our work on the vectorization of the conventional top-down traversing for simplicity. 
Despite the fact that the hybrid algorithm has shown better results
than the top-down approach, we focus our work in this paper on the
conventional top-down BFS to demonstrate how vectorization techniques
can be applied to effectively utilize Xeon Phi's vector unit. The same
techniques can be applied to the bottom-up phase, which can lead to
speed up the hybrid BFS algorithm.  Figure \ref{fig:bfslayers} shows
an example of the top-down BFS algorithm.  The exploration starts from
vertex 1 and reaches all the vertices in the three layers illustrated
by a, b and c in Figure \ref{fig:bfslayers}. Dotted lines represent
edges linked with already explored vertices.

%The first one is a set of sparse matrices from different real applications. An easy way to read these matrices is by using the Matrix Market Format [ref], which translates the graph input into a readable list of rows and columns in a text file format. The second one is the well known benchmark for graphs that contains a scalable graph \textit{Kronecker} generator that brings out graphs that naturally follow common network properties. This generator is based on the \textit{Recursive MATrix (R-MAT)} scale-free graph generation algorithm \cite{rmat}.
\subsection{The Serial Top Down BFS algorithm}
\label{sec:serBFS}
The implementation of the serial top-down BFS algorithm uses two lists
to process vertices by layers. The first list is the input list and it
contains all the vertices to be processed in the layer. The second
list is the output list which, after processing the layer will be
swapped with the input list for the next layer. The result of the
algorithm is a BFS spanning tree represented by a sequence of the
predecessors (\textit{P}) of the traversed vertices. When a vertex has
been processed it is marked as \textit{visited}, otherwise, it remains
\textit{non-visited}. Each vertex has an associated set of adjacent
vertices, known as neighbor list. Only the non-visited vertices in the
list are put into the output list to be processed in the next layer.
 
Algorithm \ref{alg:serialbfs} shows the pseudocode of the serial
top-down BFS algorithm. The pseudocode uses four data structures:
input list (\textit{in}), output list (\textit{out}), visited array
(\textit{visited}), and \textit{P}, predecessor list; all data
structures are initialized at the beginning. The visited array is used
to mark vertices as \textit{visited} during the exploration
process. Initially all the vertices are set as
\textit{non-visited}. The predecessor array is used to store the
output BFS spanning tree and is initialized with big number values,
denoted by the symbol $\infty$ in line 2. In practice, $\infty$ can be
an integer bigger than the number of vertices. The exploration starts
when the input list \textit{in} has at least one element (line
7). Thus, the starting vertex \textit{s} is placed in the input list,
visited array and in the predecessor array set as its own parent
(lines 4-6).
 
 In lines 7 to 17, every single vertex $u$ in the input list
 \textit{in} is explored. This exploration consists of checking each
 adjacent vertex $v$ of $u$ that has not been visited. If it is the
 case, then they are put into the output list \textit{out}, marked as
 \textit{visited} and the parent for the vertex $v$ in the \textit{P}
 array is set to $u$. By checking the non-visited vertices first, some
 extra work is avoided by not putting previously processed vertices in
 the list \cite{Agarwal}. In line 16, the input and the output lists
 are swapped and the output list is cleared. The algorithm ends when
 all vertices reachable from the input vertex have been marked as
 visited in \textit{out}. The output BFS spanning tree is the
 predecessors list \textit{P}, which contains a record of the order
 that the vertices were explored.
 
%One important procedure after getting the BFS output tree, is to verify its correctness. We based our BFS tree verification on the validation kernel included in the Graph500 benchmark (kernel 2). It consists on having a soft checking of the results instead of making a complete verification of the BFS tree, which would consume too much computing time.  The graph500 benchmark has its own performance metric called traversed edges per second (TEPS) in order to be able to compare results on different architectures. We aim to use this as a performance metric.

%A particular property of the BFS to point out is that vertices are discovered by layers that are in a certain \textit{distance} \textit{k} from the starting vertex \textit{s}. 

\begin{algorithm}[H]
\small
\caption{Serial \textit{Top-Down} BFS($G, s$) }
\label{alg:serialbfs}
\begin{algorithmic}[1]

\renewcommand{\algorithmicrequire}{\textbf{Initialize:}}
\REQUIRE $in.init()$
 $out.init()$
 $vis.init()$
\FORALL{ vertex $u$ $\in$ $V(G) - s$}
\STATE 
$P[u] \leftarrow  \infty$ \\
\ENDFOR
\STATE
$in.add(s)$ \\
\STATE
$vis.Set(s)$ \\
\STATE
$P[s] = s$
\WHILE{${in} \neq 0$}
\FORALL{$u \in in$} 
\FORALL{$v \in Adj[u]$} 
\IF {$vis.Test(v)$ = 0}
\STATE $vis.Set(v)$
\STATE $out.add(v)$
\STATE $P[v] = u $
\ENDIF
\ENDFOR
\ENDFOR
%\ENDFOR
\STATE $swap(in, out)$\\
 $out \leftarrow 0$
\ENDWHILE

\end{algorithmic}
\end{algorithm}

%sequential algorithm

\subsection{Parallel Top-Down BFS}
\label{sec:parBFS}
In the serial \textit{top-down} BFS algorithm, there are two levels of
parallelism that can be exploited. The first is a coarse-grain level
in the outer loop for processing the input list; the second,
finer-grain, is in the inner loop where the adjacency list is
explored. Algorithm \ref{alg:parallelbfs} shows the Parallel BFS by
augmenting Algorithm \ref{alg:serialbfs} with parallel \textit{for}
loops. In practice, we aim to parallelize the outer loop using threads and the inner loop by exploiting the vector unit. The major changes to parallelize the algorithm are in the
initialization of the lists, the visited and the predecessor array
data structures, and in the outer and inner loops, lines 8 and 9,
where all the vertices are explored in parallel.

\begin{algorithm}[H]
\small
\caption{Parallel Top-Down BFS($G, s$) }
\label{alg:parallelbfs}
\begin{algorithmic}[1]
\renewcommand{\algorithmicrequire}{\textbf{Initialize:}}
\REQUIRE $in.init()$
 $out.init()$
 $vis.init()$
\FORALL{\textbf{parallel} vertex $u$ $\in$ $V(G) - s$}
\STATE 
$P[u] = \infty$ \\
\ENDFOR
\STATE
$in.add(s)$ \\
\STATE
$vis.Set(s)$ \\
\STATE
$P[s] = s$
\WHILE{${in} \neq 0$}
\FORALL{\textbf{parallel} $u \in {in}$} 
\FORALL{\textbf{parallel} $v \in Adj[u]$} 
\IF {$vis.Test(v)$ = 0}
\STATE $vis.Set(v)$
\STATE $out.add(v)$
\STATE $P[v] = u $
\ENDIF
\ENDFOR
\ENDFOR
\STATE $swap(in, out)$\\
 $out \leftarrow 0$
\ENDWHILE
\end{algorithmic}
\end{algorithm}

\begin{figure}[h!]
\centering
\includegraphics[width=5cm]{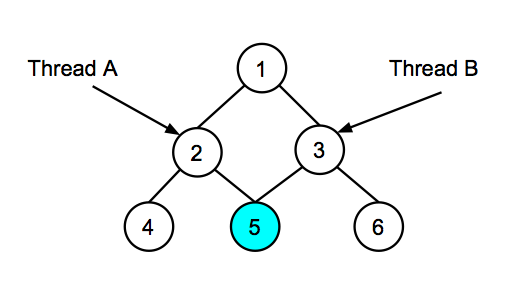}
\caption{Example of data benign race condition.}
\label{fig:race2}
\vspace{-3mm}
\end{figure}

However, this parallel version presents a race condition between
multiple threads. This condition happens in the exploration of the
adjacency list when a vertex $v$ is tested to verify if it has been
\textit{visited} previously. Multiple threads might test the same
vertex at the same time. Figure \ref{fig:race2} illustrates this race
condition. Here, threads A and B are trying to update vertex 5, which
is a child of both vertex 2 and 3. While this could end up in
redundant work when the status of the vertex and the queue are
updated, the major impact is in the predecessor list, where the parent
of vertex 5 can be set to either 2 or 3. However, this is called a
benign race condition since the correctness of the algorithm is not
affected. It means that different correct BFS spanning trees can be
generated. It is possible to avoid this race condition by using an
atomic operation such as \textit{\_\_sync\_fetch\_and\_or}. However,
we will see in the following section that another more critical race
condition comes up when we introduce bitmap arrays as data structure.

%For now, we will introduce the data structures used for our implementations. 
% as it is presented in Algorithm \ref{alg:atomicPBFS}.

%\begin{algorithm}
%\small
%\caption{Atomic Parallel \textit{Top-Down} BFS($G, s$) }
%\label{alg:atomicPBFS}
%\begin{algorithmic}[1]

%\FORALL{$u \in {in}$} 
%\FORALL{$v \in Adj[u]$} 
%\IF {$ \_\_sync\_fetch\_and\_or(vis, v) $ }
%\STATE $vis.Set(v)$
%\STATE $out.Set(v)$
%\STATE $P[u] = v $

%\ENDIF
%\ENDFOR
%\ENDFOR
%\end{algorithmic}
%\end{algorithm}

%Finally, initialisation of the involved data structures, such as: input queue, output queue, visited bitmaps and the predecessor array, needs to be parallelised. This can improve locality by distributing an initial copy of those data structures in all the cores \ref{parinitialisation}. 

%omp pragma
%synchronization 

\subsection{Parallel Top-Down BFS without bit race conditions}
\label{sub:bitrace}
%Previously, we explained the data benign race condition presented in the parallel BFS algorithm and we showed how to solve it. However, by using bitmaps, it ends up with another more critical race condition described in the following section.
In addition to the benign race condition described in Section \ref{sec:parBFS}, introducing bitmap arrays as data structures leads into a bit race condition that is solved by adding a restoration process to the parallel top-down BFS algorithm. 
\subsubsection{Data structures}

%Te falto mencionar la SCALE and edgefactor

%Our implementation uses different modules of the Graph500 benchmark
%\cite{graph500}, including the graph generator, the BFS path
%validator, the experimental design and the parallel BFS
%implementation. These details will be explained in Section
%\ref{sec:imp}. However, for now we will focus only on the data
%structures used in the BFS implementation.

\begin{figure}
\centering
\includegraphics[width=6cm]{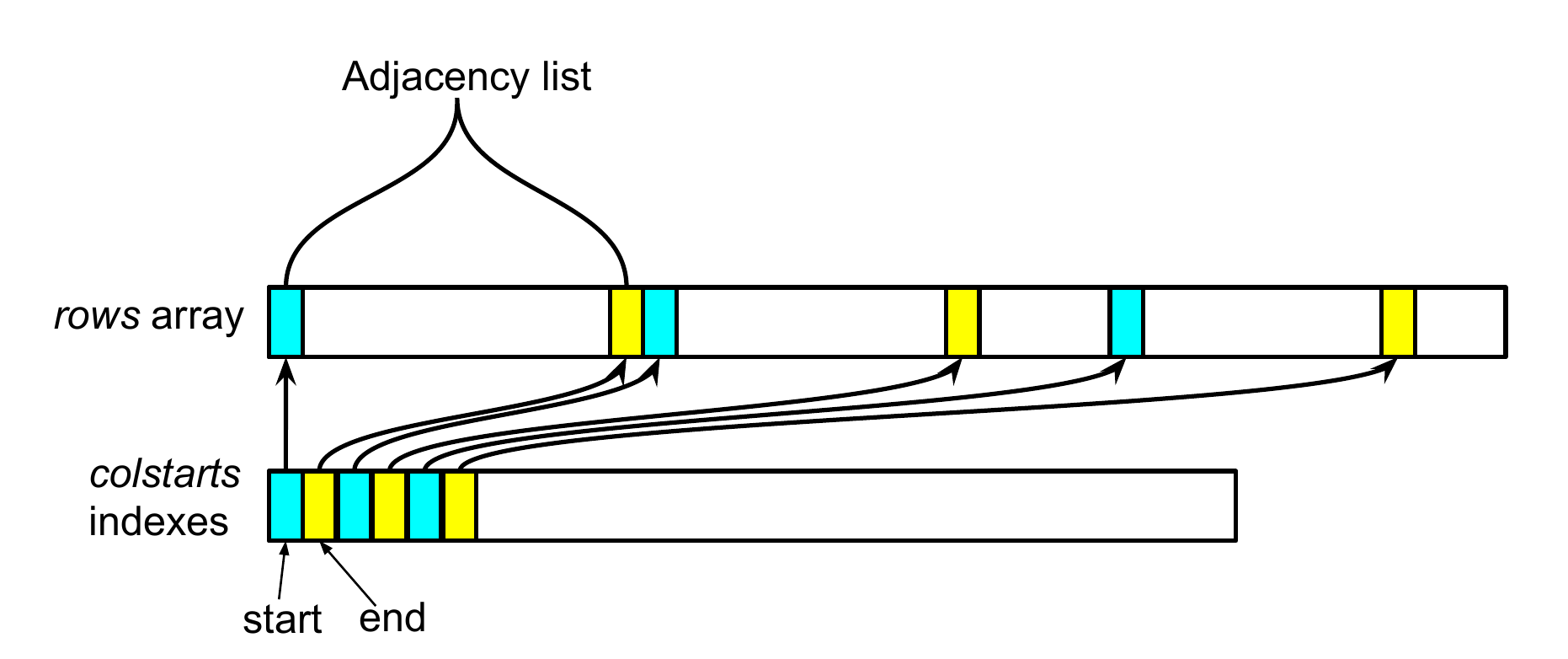}
\caption{Compressed Sparse Row representation.}
\label{fig:csr}
\vspace{-3mm}
\end{figure}

Our parallel BFS implementation is based on the implementation,
\texttt{bfs\_replicated\_csc}, given in the \textit{Graph500} source
code \cite{graph500:GIT}. The graph is efficiently represented by a
Compressed Sparse Row (CSR) matrix format, which is composed by two
integer arrays: \textit{rows} and \textit{colstarts}. The
\textit{rows} array contains the adjacency list of every vertex and
the \textit{colstarts} stores the start and the end indexes of every
vertex pointing to the \textit{rows} array. An example of this data
structure is illustrated in Figure \ref{fig:csr}. The use of the CSR is when the adjacency list is explored, line 9 in Algorithm \ref{alg:restoration} abstracts this step.

\begin{figure}
\centering
\includegraphics[width=0.5\textwidth, scale=1.0]{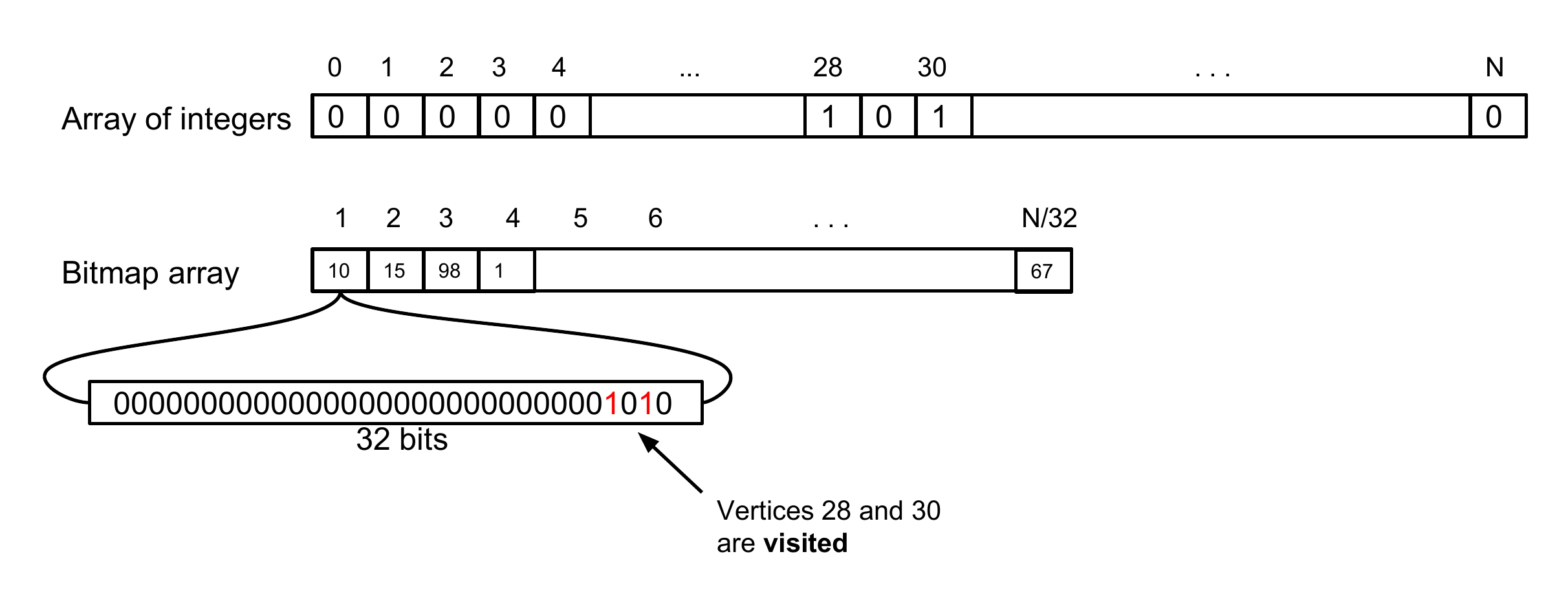}
\caption{Example of an array of integers (32 bits) represented by a bitmap array.}
\label{fig:race}
\vspace{-3mm}
\end{figure}

On the other hand, the data structures used for the input list, the
output list and the visited arrays are bitmap arrays and an integer
array for the predecessor array. A bitmap is a mapping from a set of
items to bits with values either zero or one. By having vertices
represented by a bitmap, the working set size can be reduced
significantly \cite{Agarwal}. For example, an array that holds
1,048,576 vertices represented by integers would require 4MB. By using
bitmaps this memory storage can be reduced significantly to 131,072
bytes. Figure \ref{fig:race} illustrates the mapping between an array
of integers to bits. The upper array is an array of integers, where
every vertex corresponds to each index in the array. Values can be set
to either zero or one and the length of the array is the total number
of vertices. In this example, vertex 28 and 30 are set to one. The
array at the bottom is the bitmap array and its length is the total
number of vertices divided by the length, in bits, of an integer (32
bits). Every integer in the bitmap array represents the status (0, 1)
of 32 vertices. Thus, the same vertices 28 and 30 are set to one but
they are both located in the first integer of the array as it is
illustrated.

\subsubsection{Restoration process}
\label{sec:atomic}
\begin{figure}
\centering
\includegraphics[width=5cm]{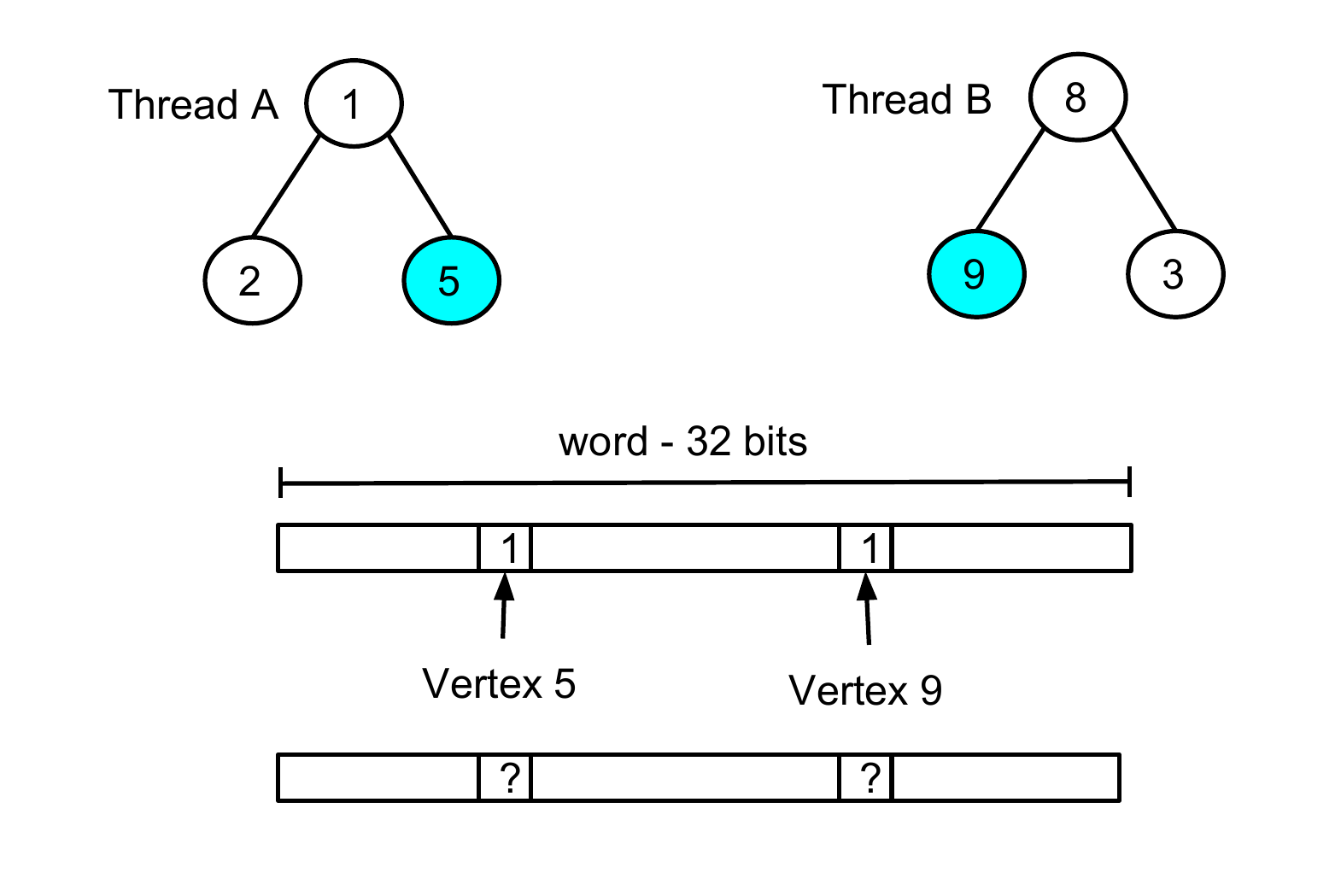}
\caption{Example of the visited bitmap array race condition.}
\label{fig:bitrace}
\vspace{-3mm}
\end{figure}

A race condition is raised by the bitmap update operations when
different threads try to update multiple bit values in the same
word. An example of this is shown in Figure \ref{fig:bitrace}, where
vertices 5 and 9 are updated by different threads but their location
in the bit array is in the same word (32 bits integer).
The bitmap race condition happens, in the adjacency list exploration, while setting the output and the visited bitmaps in lines 9 to 12 of the Algorithm \ref{alg:parallelbfs}. %but not to the predecessor list, since this one is an array of integers.
Thus, to overcome this problem a restoration process is applied as in \cite{MICgraphs} to both bitmap arrays afterwards, the visited and the output. That way, the restoration process helps to keep the output and the visited bitmaps consistent for processing the next layer. 

The restoration consists of finding all the corrupted words in the output bitmap array that were updated, if a bit race condition happened the output array should have at least one bit set. Since the predecessor list is an array of integers, it does not present the bit race condition and is used to fix the corrupted output array. To do so, first we identify which vertices were updated in the predecessor list by setting a negative number, which is a subtraction of the number of nodes from the vertex's parent. Second, we iterate through the non zero words (32-bits integer) in the output bitmap array. In those words are the vertices that are corrupted and need to be fixed. We step through each of the 32 bits in the word to look for the corresponding vertices, in lines 20 and 21 of Algorithm \ref{alg:restoration}, that have been set to a negative number in the predecessor list. These vertices are restored by setting their corresponding bit in the output bitmap and adding back the number of nodes in the predecessor list. Finally, the visited bitmap array is updated consistently. 

%The restoration consists of finding all the non-zero words in the output queue to compare them with the predecessor array. To identify the vertices that need to be restored, the P array is set to a negative number by substrating the number of nodes to the vertex's parent. The 32 bits of each non-zero word are tested against the \textit{P} array. Thus, if the respective vertex, taken from the 32 bits iteration, has been set to a negative number in the predecessor array, this vertex needs to be restored. The output queue and the visited bitmaps are updated in the specific vertex location and the P array is set to a positive number. 
%An optimization was introduced in line X by testing if the vertex is not either in the visited array nor the output queue. Also, the visited array is set until the restoration since is when the vertices are consistent. 

Although the restoration process implies extra work, it solves the bit race condition without having to use atomic operations while keeping correctness and even more important allow us to use it in the vectorization process described in Section \ref{sec:vectorization}. Algorithm \ref{alg:restoration} shows the complete pseudocode of the BFS algorithm containing the restoration process to cope with the bitmap race condition, lines 15 to 30. However, the benign race condition described in \ref{sec:parBFS} still remains since we avoid making use of atomic operations for further vectorization. Bitmap operations used by the visited, input and output arrays are: $InitBitmap()$, $SetBit(n)$, $GetBit(n)$, $TestBit(n)$ and $bit2vertex(n)$, where $n$ is the bit position.

\begin{algorithm}
\small
\caption{Parallel BFS without bit race conditions.}
\label{alg:restoration}
\begin{algorithmic}[1]

\renewcommand{\algorithmicrequire}{\textbf{Initialize:}}
\REQUIRE $in.InitBitmap()$
 $out.InitBitmap()$
 $vis.InitBitmap()$
\FORALL{\textbf{parallel} vertex $u$ $\in$ $V(G) - s$}
\STATE 
$P[u] = \infty$ \\
\ENDFOR
\STATE
$in.SetBit(s)$ \\
\STATE
$vis.SetBit(s)$ \\
\STATE
$P[s] = s$

\WHILE{${in} \neq 0$}
\FORALL{\textbf{parallel} $u \in {in}$} 
\FORALL{\textbf{parallel} $v \in Adj[u]$} 
\IF {$v \notin ($$vis.TestBit(v)$ OR $out.TestBit(v) )$ }
\STATE $out.SetBit(v)$
\STATE $P[v] = u - nodes$
\ENDIF
\ENDFOR
\STATE //Restoration process
\FORALL{\textbf{parallel} $w \in {out}$} 
\STATE // w is a word in out bitmap 
\IF {$w \neq 0 $}
	\STATE // iterate through every bit in w
	\FORALL{ $b \in w$}
		\STATE vertex = bit2vertex(b) 
		\IF {$P[vertex] < 0$}
			\STATE $out.SetBit(vertex)$
			\STATE $vis.SetBit(vertex)$
			\STATE $P[vertex] = P[vertex] + nodes $
		\ENDIF
	\ENDFOR
\ENDIF
\ENDFOR
\ENDFOR
\STATE $swap(in, out)$\\
 $out \leftarrow 0$
\ENDWHILE
\end{algorithmic}
\end{algorithm}

\section{BFS Vectorization}
\label{sec:vectorization}
%why are you vectorizing the exploration of the neighbours?

Our vectorized BFS algorithm is based on the parallel BFS algorithm without using atomic operations presented in Algorithm \ref{alg:restoration}. Basically, it avoids atomic operations by using an extra step to restore possible missing values in the output queue due to a data race condition and the lack of atomic operations at bit level. This atomicity freedom allows to vectorize the algorithm straight away because atomic bit operations are not part of the instruction set architecture (ISA) in the Intel compiler\cite{xeonInstSet}.

\begin{figure}[h!]
\centering
\includegraphics[width=0.5\textwidth, scale=0.5]{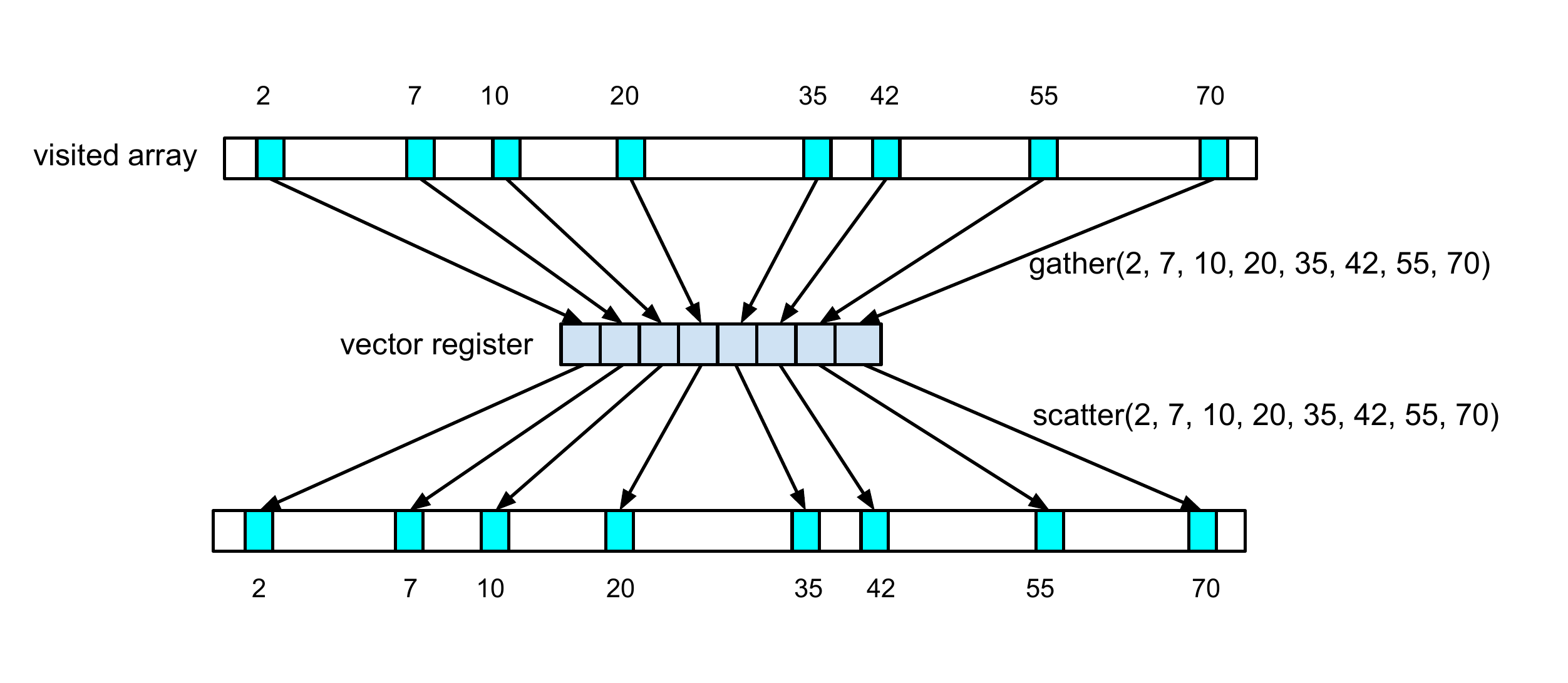}
\caption{\textit{Gather} and \textit{scatter} operations, to load non-contiguous visited array.}
\label{fig:gather_scatter}
\vspace{-3mm}
\end{figure}

\begin{figure}
\centering
\includegraphics[width=0.5\textwidth, scale=0.5]{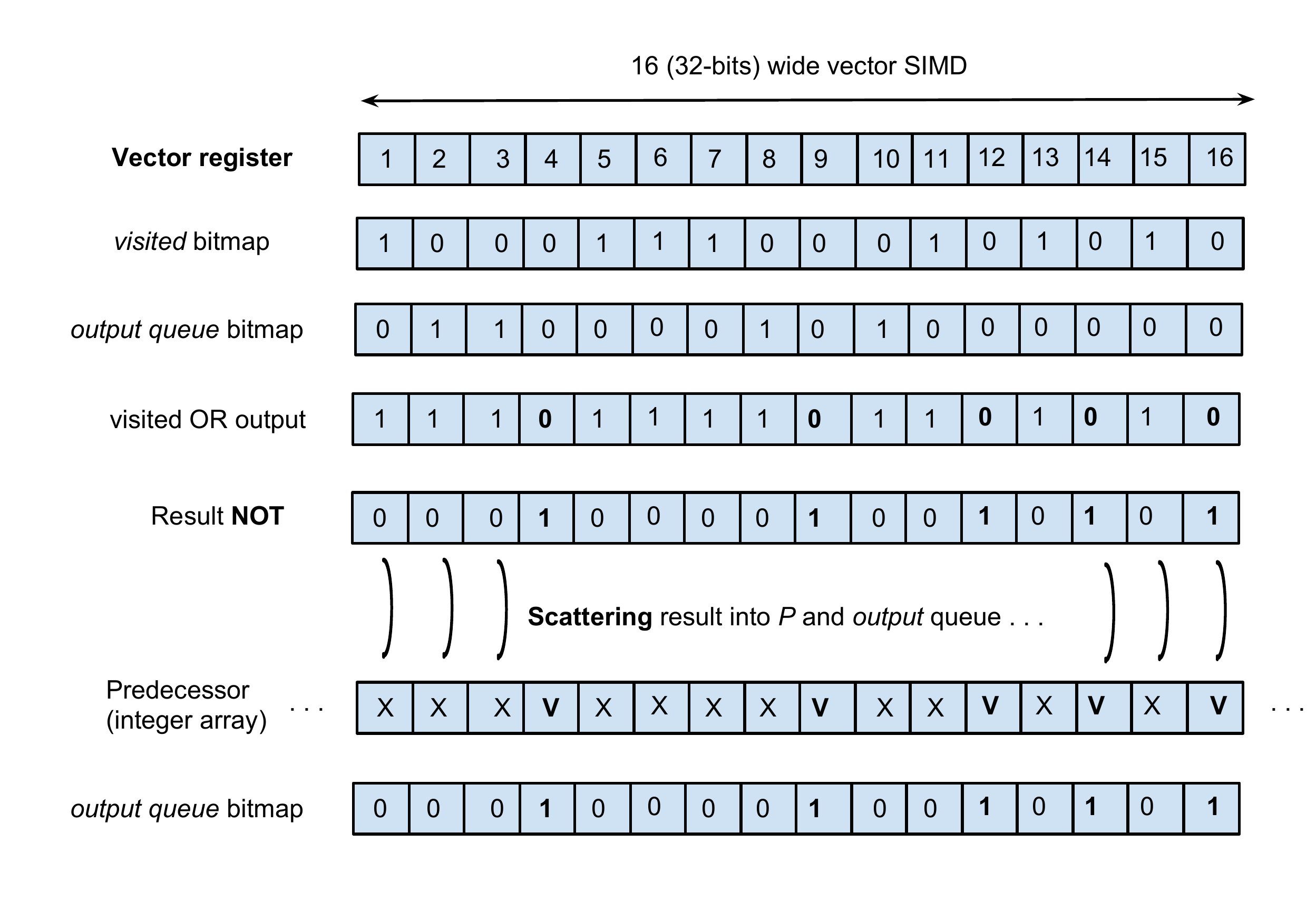}
\caption{Example of vectorizing the adjacency list exploration.}
\label{fig:simdlane}
\vspace{-3mm}
\end{figure}

There are two potential parts to be vectorized in the algorithm: the adjacency list exploration and the restoration process. In the adjacency list exploration, each element in the list is explored sequentially. By using the vector unit, instead of exploring one element at a time, it could be possible for one thread, to explore 16 (32-bits integers) vertices simultaneously. The vectorization of the adjacency list exploration involves three main SIMD steps. Firstly, a sequence of vertices in the adjacency list are loaded into the vector unit, which can hold 16 (32-bits) vertices. Secondly, all the loaded vertices are filtered by using the visited array and the output queue bitmaps to find the ones that have not been visited yet either in previous layers (visited array) or in the current layer (output queue). 
%Unlike the serial BFS algorithm \ref{},  the filtering condition includes the output queue \textit{out}.
% According with  \cite{Agarwal}, this is done to reduce the atomic operations by setting \textit{vis} at last and union it with the output queue \textit{out}.  
Finally, the result is set back to the predecessor array \textit{P} and the output queue \textit{out}. Figure \ref{fig:simdlane} shows an example of the vectorization of the adjacency list exploration in a vector register of 512 bits wide. The specific values in the visited and output queue bitmap arrays are loaded into the vector unit by using the SIMD \textit{gather} instructions. The \textit{scatter} and the \textit{gather} operations are two instructions that the Xeon Phi use to deal with non-contiguous memory loads to the vector register and data stores back to memory. Both operations receive, as argument, a list of indexes to be scattered/gathered in the array. Figure \ref{fig:gather_scatter} illustrates both instructions to update the visited array. Since the visited array is a bitmap array an index transformation is necessary to load the bit word (32 bits) specific to a vertex. Then, the vector unit can do bit shifting operations to get the bit value of the vertex.

Once the specific values of the visited array and the output queue are loaded into the vector unit, a filtering process is applied. To find all the vertices that have not been either visited previously (upper layers) or put into the output queue recently (current layer), two logical operations (OR and NOT) are used to create a vector mask. The \textit{OR} operation will find the union of the vertices that have already been visited and the ones that have been put into the output queue. By applying the \textit{NOT} logical operation, we can find all the vertices that have not been visited or set into the output queue. Afterwards, the result is scattered into the predecessor array and the output queue, only those indexes that have one as bit value in the mask are updated. 

On the other hand, the vectorization of the restoration process adds an extra step to the parallel Algorithm \ref{alg:restoration} and consists of repairing the output queue and the visited array based on the P array, which remains consistent. To vectorize it, there are some details to take into account. Firstly, there is a difference between the output queue and the \textit{P} array representation. While the output queue is a bitmap array, \textit{P} is an array of integers. Thus, during stepping through each of the words (32 bits length) of the output queue, 32 vertices are expected to be processed, however the vector unit can only process up to 16 elements at a time. To cope with that, we split the word in two: the low part and the high part. Then, the restoration process is divided into two sections, the first one repairs the vertices that are located in the low part and the second section repairs vertices located in the high part. Source code of the SIMD implementation is presented at the end of Section \ref{sub:xeonphi}. 
% \ref{sub:xeonphi} shows the source code of the vectorization of the restoration process. 

%\begin{figure}[h!]
%\centering
%\includegraphics[width=\textwidth, scale=0.5]{./img/restoration.png}
%\caption{SIMD vectorization of the BFS atomic free restoration process.}
%\label{fig:simdrestoration}
%\end{figure}
%\begin{figure*}[h!]
%\centering
%\includegraphics[width=0.8\textwidth, scale=0.5]{./img/restoration.png}
%\caption{SIMD vectorization of the BFS atomic free restoration process.}
%\label{fig:simdrestoration}
%\end{figure*}

\subsection{Which layers are vectorized?}
Large and sparse graphs often present the \textit{small-world} graphs
properties of having small diameter and skewed degree distribution
\cite{RMAT}. We use RMAT, a synthetic graph generator, to create our
input \textit{small-world} large graphs. The RMAT graph size is
defined by two input values: the \textit{SCALE} and the
\textit{edgefactor}. The total of vertices in the graph is calculated
by $2^{SCALE}$ and the number of edges generated by $2^{SCALE} *
edgefactor$, including self-loops and repeated edges. Graph structure
is crucial because it can help to improve performance by exploiting
the architecture resources efficiently, such as the vector unit on the
Xeon Phi. Table \ref{tab:vectorization} shows the number of input
vertices, the number of edges and the traversed vertices by the BFS
top-down algorithm per layer, for a graph of 1,048,576 million
vertices (RMAT graph with \textit{SCALE} 20 and edgefactor 16),
choosing the starting vertex randomly. As it can be seen, the number
of input vertices per layer increase along with the number of layer
until the middle layer is reached and then they start to decrease. The
diameter of the graph is 7, which is reflected in having 7 traversed
layers. On the other hand, the number of edges per layer varies
according to the \textit{edgefactor}, which is used to distribute the
edges per vertex. Both graph characteristics, diameter and vertex
degree are key to decide what is the best way to improve threads
workload imbalance and to increase the vector unit usage. Assuming
that most of the vertices are traversed in the first layers, we used
the vectorized SIMD BFS top-down algorithm only for the first two
layers and the parallel top-down presented in Algorithm
\ref{alg:parallelbfs} for the rest of the layers. 

\begin{table}
\centering
\caption{Traversed vertices per layer for a 1,048,575 million vertices graph, \textit{SCALE} 20 and \textit{edgefactor} 16.}
\label{tab:vectorization}
 \begin{tabular}{|r|r|r|r|r|} 
 \hline
 \textbf{Layer} & \textbf{Vertices}  & \textbf{Edges} & \textbf{Traversed vertices} \\ 
 \hline
0 & 1	&12 & 12\\
1 & 12 &	21,892& 18,122\\
2 & 18,122	&13,547,462& 540,575\\
3 & 540,575 &	17,626,910 & 100,874\\
4 & 100,874 & 150,698 & 486\\
5 & 486	& 490 & 4\\
6 & 2 & 2 & 0 \\
 \hline
\end{tabular}
\vspace{-3mm}
\end{table}

%number of vertices in the frontier
%1.- Explicar el comportamiento y el paralelismo
%2.- vector usage rate
%3.- threads load imbalance
%4.- Implementing vectorization in the middle layers helps to mitigate low vector usage overhead.
%5.-  

%//Presentar Tabla del 
%Numero de vertices en la frontier
%Numero promedio de neighbours per node in the frontier - vector usage rate
%Explicar que el numero de theads se exploran contando el numero de nodos en la frontier o en cada layer. 
%to point out that others factor are involved specific for the Xeon Phi architecture such as thread affinity 
%236 cores - >> specially when different threads are sharing the vector unit, so when all full cores are used. 

\subsection{Xeon Phi optimizations}
\label{sub:xeonphi}
To optimize the vectorization of the BFS top-down algorithm, there are some crucial factors such as cache-oriented memory address alignment, loop vectorization, masking and prefetching \cite{vecguide}. 

%automatic vectorization
%manual vectorization
\paragraph{\textbf{Intrinsic functions}}
Despite the promise of automatic vectorization, this is not
straightforward for algorithms with irregular data access patterns,
such as the BFS. Some explicit manual code transformations are needed
to assist the compiler. An explicit way to use the vector unit is
through \textit{intrinsic functions}, these are a set of assembly
functions that allow complete control over the vector unit. We used
specialized intrinsic functions which are part of the Intel AVX-512
instruction set.
\paragraph{\textbf{Data alignment}} 
Data alignment specifications assist the compiler operation in the
creation of objects on specific byte boundaries
\cite{xeonphiBook}. The optimal data alignment for the Xeon Phi is on
64 byte boundaries. To align the \textit{rows} array, we used the
intrinsic function \textit{mm\_malloc}. This data alignment allows the
vector unit to have more efficient access to the memory. However, due
to the way the \textit{rows} array is built, it might lead to some
\textit{less-than-full-vector} loop vectorization inefficiencies, such
as the \textit{peel} and the \textit{remainder} loops
\cite{practicalSIMD}, caused non-aligned boundary accesses.
\paragraph{\textbf{Peel and remainder loops}}
The \textit{peel} loop is the sequence of contiguous elements for
which the start index in the array does not match with a aligned
boundary of the array. On the contrary, the remainder loop is the
sequence of the last elements, the tail, that do not fit on an aligned
boundary. Both cases imply an extra processing step because they
cannot be computed as a complete 16 elements vector. There are
different approaches to cope with the implications of having
\textit{less-than-full-vector} loops including \textit{padding},
sequential processing of \textit{peel} and \textit{remainder} loops,
and the use of vector masks.

\paragraph{\textbf{Prefetching}}
\label{par:prefet}
%This technique is important to improve performance so it is implemented in several commercial architectures such as the Xeon Phi. The Xeon Phi %contains both hardware (HWP) and software (SWP) prefetching. By using the Intel compiler flag optimization (-O2 or higher), by default hardware prefetching is activated. In such case, the compiler guides the hardware to fetch data that is more likely to be referenced before it is actually needed. 
Prefetching is a technique that helps hide memory latency. The Xeon
Phi has a heuristic to choose between hardware or software
prefetching. Due to the irregular data access patterns that the BFS
algorithm shows, software prefetching is necessary. In addition,
working with large graph sizes might result in having large numbers of
L2 cache misses, and thus poor performance. Prefetch {\em distance} is
a metric that hints to the compiler the right number of cycles to load
data ahead of it use. Finding the right distance is crucial to gain
performance \cite{Mehta:2014}. A recommended distance should be one
similar to memory latency \cite{Badawy04}. Rather than calculate a
precise distance, and based on the work in \cite{JhaHLCH15}, in the
adjacency list exploration we prefetch the rows array for the vertices
that will be processed in the next iteration. Also, we use prefetching
intrinsic functions for every data load/store to the vector unit by
setting a hint that indicates which memory will be prefetched to
either L1 cache (\texttt{\_MM\_HINT\_T0}) or to L2
cache(\texttt{\_MM\_HINT\_T1}).
%\vspace{2mm}

The vectorization of the adjacency list exploration consists of
splitting the list into chunks of 16 (32 bit) elements. The peel and
the remainder loops are considered special cases because vertices need
to be filtered according to the pre-calculated mask. Listing
\ref{listing:SIMDFlt} shows the source code of the vectorization of
the adjacency list for a \textit{full-vector}, using optimizations
such as alignment, masks and prefetching. The code consists of three
main steps described previously in the algorithm: loading the
adjacency list, filtering non-visited vertices and setting the values
back to memory. However, an intermediate operation is needed due to
the discrepancy between the indexes of the 16 input vertex list
(32-bit integers) and the bitmap arrays (bits) in the step 2:
filtering the unvisited vertices in the visited array and the output
queue. This step is implemented by getting the word and the bit offset
of each element in the adjacency list. The words vector is used to
gather all the words to be updated from the visited array and the
output queue. The bit offset vector is used to create a mask to filter
the specific bit values by shifting it to the left. The words and the
bit offsets are used to generate a vector mask that allows to filter
the visited and the output queue bitmap arrays respectively. 
%//https://bitbucket.org/Mireya/bfs}.
\begin{lstlisting}[caption=Adjacency list exploration using SIMD intrinsic functions.,
                   label=listing:SIMDFlt]
/* 1.- Load adjacency list to the register */
__m512i vneig = _mm512_load_epi32(&rows[index * 16);
	
/* Getting word and bit offset */
__m512i vword = _mm512_div_epi32(vneig, _mm512_set1_epi32(BITS_PER_WORD));
__m512i vbits = _mm512_rem_epi32(vneig, _mm512_set1_epi32(BITS_PER_WORD));
	
/* Gathering words from visited bitmap array */
_mm512_prefetch_i32gather_ps(vword, explored->start, sizeof(word_t), _MM_HINT_T0);
_mm512_prefetch_i32gather_ps(vword, queue->start, sizeof(word_t), _MM_HINT_T0);

__m512i vis_words = _mm512_i32gather_epi32(vword, explored->start, sizeof(word_t));
__m512i out_words = _mm512_i32gather_epi32(vword, queue->start, sizeof(word_t));
	
/* Shifting 1 to the left indexes position in the vneig array */	
__m512i bits = _mm512_sllv_epi32(_mm512_set1_epi32(1), vbits);

__mask16 mask = _mm512_knot(_mm512_kor(_mm512_test_epi32_mask(vis_words, bits), _mm512_test_epi32_mask(out_words, bits)));

/*3.- Scattering P (bfs_tree) and output queue */
_mm512_mask_prefetch_i32scatter_ps(bfs_tree, mask, vneig, sizeof(word_t), _MM_HINT_T0);
_mm512_mask_i32scatter_epi32(bfs_tree, mask, vneig, _mm512_set1_epi32(vertex - nodes), sizeof(word_t));

/* Setting the output queue */
//Adding to the output queue word the new bit values depending on the filtered mask. 
__m512i new_values = _mm512_mask_or_epi32(_mm512_set1_epi32(0), mask, out_words, bits);

_mm512_mask_prefetch_i32scatter_ps(queue->start, mask, vword, sizeof(word_t), _MM_HINT_T0);
_mm512_mask_i32scatter_epi32(queue->start, mask, vword, new_values, sizeof(word_t));

\end{lstlisting}

\begin{figure}
\centering
\includegraphics[trim = 10mm 24mm 23mm 21mm, clip, width=0.5\textwidth]{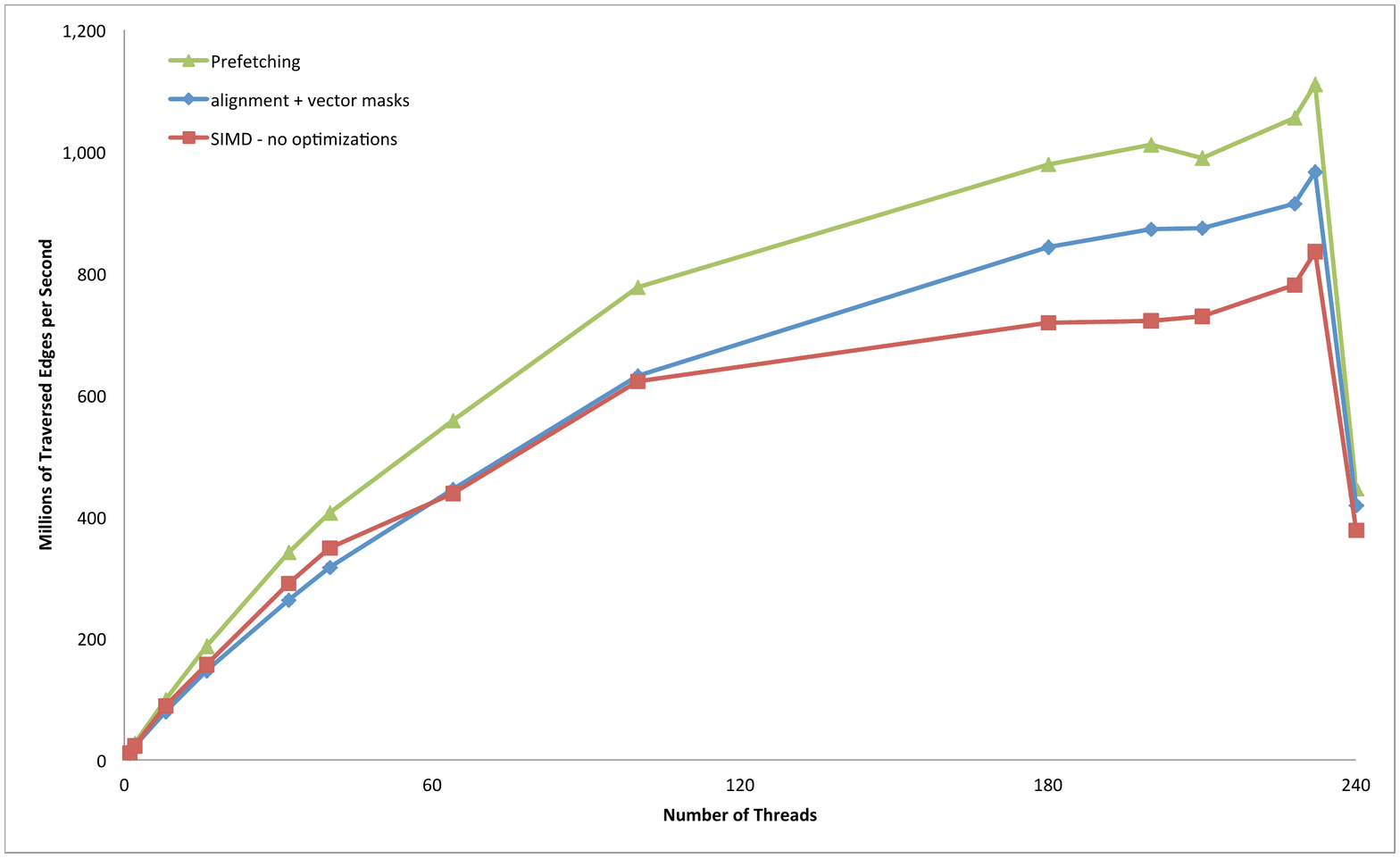}
\caption{BFS experimental optimizations results for a graph of \textit{SCALE} 20 and \textit{edgefactor}=16.}
\label{fig:noopt}
\vskip -6pt
\end{figure}

 Figure \ref{fig:noopt} shows different results for the BFS top-down
 native algorithm on the Xeon Phi. The experiments involved three
 implementations, including the SIMD without optimizations (SIMD - no
 opt), the combined (SIMD + parallel) BFS algorithm plus alignment and
 masks optimizations and the one with prefetching. As it can be seen
 performance was increased after applying prefetching. Future work
 will target improving prefetching by finding a good prefetch
 distance.
 
%\begin{figure*}[h!]
%\centering
%\includegraphics[width=1.0\textwidth, scale=0.5]{./img/code.png}
%\caption{ Vectorization implementation using intrinsic functions.}
%\label{fig:intrinsics}
%\end{figure*}
\paragraph{\textbf{Thread affinity}}
Thread affinity is a feature in hyper-threading architectures
that enables the pinning of threads to specific {\em logical cores} on
the same physical core, enabling the user to affect the use of shared
resources in a core (e.g. cache and memory bandwidth). The Xeon Phi
allows up to 4 hardware threads per core and provides three strategies
for controlling thread affinity: compact, scatter and balanced. {\em
  Compact} affinity assigns free threads as close as possible to the
thread contexts in a core (i.e. to logical cores on the same physical
core); {\em scatter} distributes the threads as widely as possible
across the entire set of physical cores, using one logical core per
physical core for each thread placed up to the number of physical cores
and then cycling through the physical cores again. Finally, {\em
  balanced} is similar to scatter but places threads with adjacent
thread ids on the same core. Regardless of the affinity strategy used,
if the number of threads is equal to the number of logical cores
available, the affinity is equivalent to compact (i.e. four threads
per physical core) and resource sharing is at a maximum.

Since the effect of thread affinity on performance is
application dependant (for example, sharing cached data between thread
may be beneficial but sharing memory bandwidth may not, depending on
the application), we ran some experiments to determine which strategy
works best for our BFS algorithm. By experimenting we found that
balanced affinity was generally better and it was used in the results
presented in Figure \ref{fig:noopt}. However, we followed the methodology presented in \cite{RodchenkoNPL15} to demonstrate the
effect of affinity on the BFS algorithm, we ran a 48 thread version
manually controlling the affinity to achieve one, two, three and four
threads per core (1T/core, 2T/core, 3T/core and 4T/core), thus using 48
cores with one thread per core down to 12 cores with four threads per
core. The thread affinity is controlled through the environment
variable \textit{KMP\_AFFINITY}.
\clearpage
\section{Experimental Setup}
\label{sec:exp}
%Te falto mencionar la SCALE and edgefactor
\subsection{Hardware platform}
We evaluate the \textit{non-SIMD} version presented in Algorithm
\ref{alg:parallelbfs} and the vectorized version, \textit{simd}, of
the BFS top-down algorithm on the Intel Xeon Phi. We used OpenMP as a
thread parallel platform and Linux as execution platform~\footnote{The Linux operating system allows access to the OpenMP library with flag \textit{-fopenmp}.}. We compiled our code with the Intel C++
compiler (version 14.0.0) with the optimization flag \textit{-02} and
\textit{-fopenmp}~\footnote{Higher optimisation levels did not improve
  performance}. We used the intrinsic functions which give access to
the AVX-512 instructions set. We used Linux as platform.

\subsection{Input graph}
Our implementation uses different modules of the Graph500 benchmark
\cite{graph500}, including the graph generator, the BFS path
validator, the experimental design and the ability to run our parallel
BFS implementation.  Firstly, the graph generator creates synthetic
scalable large \textit{Kronecker} graphs \cite{Kronecker} and is based
on the R-MAT random graph model \cite{RMAT}. These graphs aim to
naturally generate graphs with common real network properties in order
to be able to analyse them. The graph size is defined by the
\textit{SCALE} and the \textit{edgefactor} values. The total number of
vertices in the graph is calculated by $2^{SCALE}$ and the number of
edges generated by $2^{SCALE} * edgefactor$ * 2 (the factor of 2
reflects the fact that the edges are bidirectional). The generator
uses four initiator parameters (\textit{A, B, C} and \textit{D}),
which are the probabilities which determine how edges are distributed
in the adjacency matrix representing the graph. We used the standard
set of parameters defined by Graph500 (A=0.57, B=0.19, C=0.19 and
D=0.05).

\subsection{Implementation details}
\label{sec:imp}
The input graph is efficiently represented by a Compressed Sparse Row
(CSR) matrix format. The validation method, which checks the
correctness of the algorithm, consists of five check results that do
not intend to get a full check of the generated output (the bfs spanning tree) but just provide a `soft' check of the output.  The
experimental design comprises 64 BFS executions each with a randomly
chosen different starting vertex. The time of each execution is
measured in seconds. After the completion of the executions,
statistics, including time and Traversed Edges Per Second (TEPS), are
collected. TEPS is a \textit{Graph500} performance metric used to
compare other BFS implementations on different architectures. However,
it became apparent that out of the 64 BFS iterations, some of the
starting points are unconnected, resulting in zero TEPS values for
those iterations. This then results in having a harmonic mean, as
calculated by Graph500, higher than the maximum number of TEPS. Some
groups filter out such unconnected nodes in their experiments. Our
results show the harmonic mean of the TEPS across the 64 executions
without filtering in order to compare fairly with other
implementations such as \cite{MICgraphs} and \cite{Beamer:2012}.
 
{\bf Result presentation:} The experimental results reported result
from a sequence of sets of 64 executions (one for each selected start
vertex) in which we varying the following parameters: the number of
threads and the graph
\textit{SCALE} factor (the edgefactor is fixed at 16). The number of threads were chosen as: 1, 2,
8, 16, 32, 40, 64, 100, 180, 200, 210, 228, 232 and 240 (the maximum
number of one thread per logical core). The \textit{SCALE} were set to
18, 19 and 20.

%This
                                                                                                                                                                                                                                                                                                                                                                                                                                                                                                                                                                                                                                                                                                                                                  %variable
                                                                                                                                                                                                                                                                                                                                                                                                                                                                                                                                                                                                                                                                                                                                                  %can
                                                                                                                                                                                                                                                                                                                                                                                                                                                                                                                                                                                                                                                                                                                                                  %set
                                                                                                                                                                                                                                                                                                                                                                                                                                                                                                                                                                                                                                                                                                                                                  %to
                                                                                                                                                                                                                                                                                                                                                                                                                                                                                                                                                                                                                                                                                                                                                  %automatic
                                                                                                                                                                                                                                                                                                                                                                                                                                                                                                                                                                                                                                                                                                                                                  %layouts
                                                                                                                                                                                                                                                                                                                                                                                                                                                                                                                                                                                                                                                                                                                                                  %such
                                                                                                                                                                                                                                                                                                                                                                                                                                                                                                                                                                                                                                                                                                                                                  %as:
                                                                                                                                                                                                                                                                                                                                                                                                                                                                                                                                                                                                                                                                                                                                                  %balanced. compact
                                                                                                                                                                                                                                                                                                                                                                                                                                                                                                                                                                                                                                                                                                                                                  %and
                                                                                                                                                                                                                                                                                                                                                                                                                                                                                                                                                                                                                                                                                                                                                  %scatter. However,
                                                                                                                                                                                                                                                                                                                                                                                                                                                                                                                                                                                                                                                                                                                                                  %in
                                                                                                                                                                                                                                                                                                                                                                                                                                                                                                                                                                                                                                                                                                                                                  %this
                                                                                                                                                                                                                                                                                                                                                                                                                                                                                                                                                                                                                                                                                                                                                  %experiment,
                                                                                                                                                                                                                                                                                                                                                                                                                                                                                                                                                                                                                                                                                                                                                  %we
                                                                                                                                                                                                                                                                                                                                                                                                                                                                                                                                                                                                                                                                                                                                                  %explored
                                                                                                                                                                                                                                                                                                                                                                                                                                                                                                                                                                                                                                                                                                                                                  %thread-core
                                                                                                                                                                                                                                                                                                                                                                                                                                                                                                                                                                                                                                                                                                                                                  %affinity
                                                                                                                                                                                                                                                                                                                                                                                                                                                                                                                                                                                                                                                                                                                                                  %mapping
                                                                                                                                                                                                                                                                                                                                                                                                                                                                                                                                                                                                                                                                                                                                                  %for
                                                                                                                                                                                                                                                                                                                                                                                                                                                                                                                                                                                                                                                                                                                                                  %a
                                                                                                                                                                                                                                                                                                                                                                                                                                                                                                                                                                                                                                                                                                                                                  %fixed
                                                                                                                                                                                                                                                                                                                                                                                                                                                                                                                                                                                                                                                                                                                                                  %number
                                                                                                                                                                                                                                                                                                                                                                                                                                                                                                                                                                                                                                                                                                                                                  %of
                                                                                                                                                                                                                                                                                                                                                                                                                                                                                                                                                                                                                                                                                                                                                  %threads. We
                                                                                                                                                                                                                                                                                                                                                                                                                                                                                                                                                                                                                                                                                                                                                  %ran
                                                                                                                                                                                                                                                                                                                                                                                                                                                                                                                                                                                                                                                                                                                                                  %different
                                                                                                                                                                                                                                                                                                                                                                                                                                                                                                                                                                                                                                                                                                                                                  %executions
                                                                                                                                                                                                                                                                                                                                                                                                                                                                                                                                                                                                                                                                                                                                                  %for
                                                                                                                                                                                                                                                                                                                                                                                                                                                                                                                                                                                                                                                                                                                                                  %48
                                                                                                                                                                                                                                                                                                                                                                                                                                                                                                                                                                                                                                                                                                                                                  %number
                                                                                                                                                                                                                                                                                                                                                                                                                                                                                                                                                                                                                                                                                                                                                  %of
                                                                                                                                                                                                                                                                                                                                                                                                                                                                                                                                                                                                                                                                                                                                                  %cores,
                                                                                                                                                                                                                                                                                                                                                                                                                                                                                                                                                                                                                                                                                                                                                  %since
                                                                                                                                                                                                                                                                                                                                                                                                                                                                                                                                                                                                                                                                                                                                                  %it
                                                                                                                                                                                                                                                                                                                                                                                                                                                                                                                                                                                                                                                                                                                                                  %is
                                                                                                                                                                                                                                                                                                                                                                                                                                                                                                                                                                                                                                                                                                                                                  %a
                                                                                                                                                                                                                                                                                                                                                                                                                                                                                                                                                                                                                                                                                                                                                  %multiple
                                                                                                                                                                                                                                                                                                                                                                                                                                                                                                                                                                                                                                                                                                                                                  %of
                                                                                                                                                                                                                                                                                                                                                                                                                                                                                                                                                                                                                                                                                                                                                  %4
                                                                                                                                                                                                                                                                                                                                                                                                                                                                                                                                                                                                                                                                                                                                                  %it
                                                                                                                                                                                                                                                                                                                                                                                                                                                                                                                                                                                                                                                                                                                                                  %helps
                                                                                                                                                                                                                                                                                                                                                                                                                                                                                                                                                                                                                                                                                                                                                  %to
                                                                                                                                                                                                                                                                                                                                                                                                                                                                                                                                                                                                                                                                                                                                                  %balance
                                                                                                                                                                                                                                                                                                                                                                                                                                                                                                                                                                                                                                                                                                                                                  %the
                                                                                                                                                                                                                                                                                                                                                                                                                                                                                                                                                                                                                                                                                                                                                  %4
                                                                                                                                                                                                                                                                                                                                                                                                                                                                                                                                                                                                                                                                                                                                                  %threads
                                                                                                                                                                                                                                                                                                                                                                                                                                                                                                                                                                                                                                                                                                                                                  %workload
                                                                                                                                                                                                                                                                                                                                                                                                                                                                                                                                                                                                                                                                                                                                                  %equitably.

%\begin{figure}[h!]

%\centering
%\includegraphics[width=0.5\textwidth]{./img/results/test.pdf}
%\caption{Experiment results from the non-simd version.}
%\label{fig:nosimd}
%\end{figure}

%\begin{figure}[h!]
%\centering
%\includegraphics[width=0.5\textwidth]{./img/results/simd.png}
%\caption{The Intel \textsuperscript{\textregistered} Xeon Phi Microarchitecture, figure taken from \cite{MICgraphs}.}

%\label{fig:simd}
%\end{figure}

\section{Results and Analysis}
\label{sec:res}
\subsection{SIMD version versus non-SIMD version}
\begin{figure}
    \begin{subfigure}{0.3\textwidth}
        \centering
        \includegraphics[width=90mm,scale=0.2]{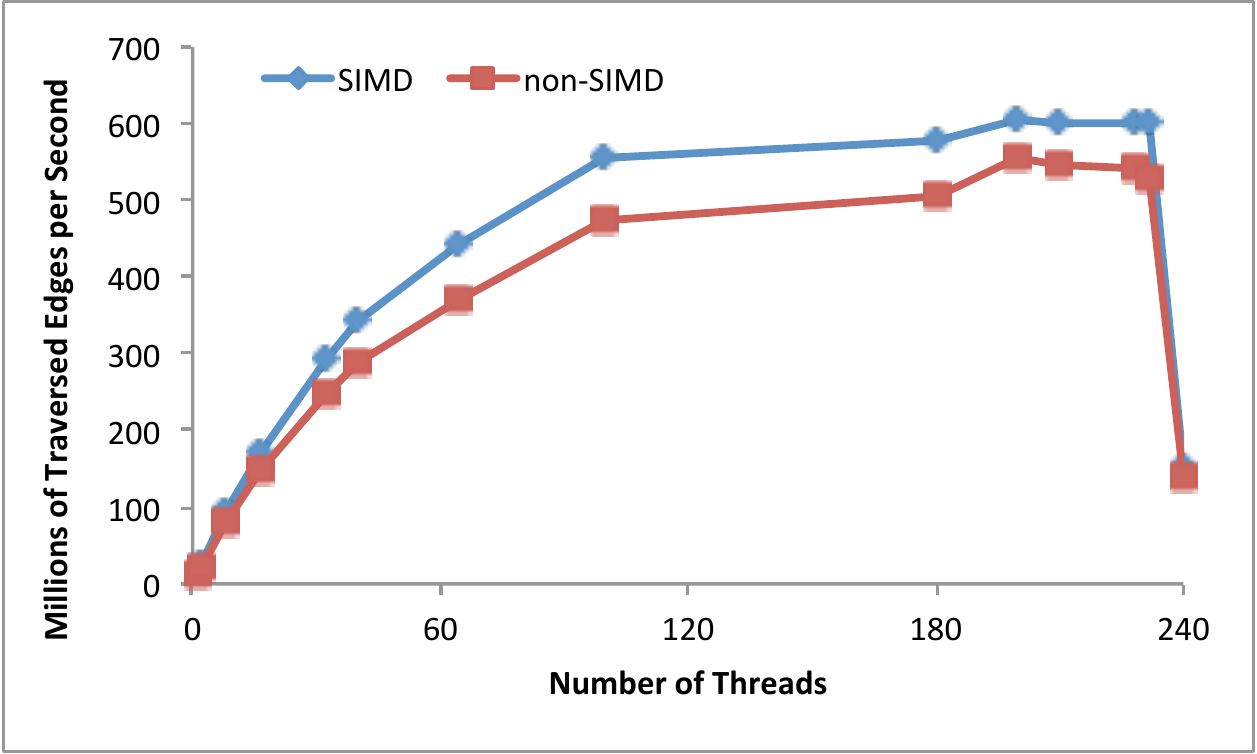}
      \caption{$SCALE=18$ and $edgefactor=16$.}
        \label{fidatag:nosimd}
    \end{subfigure}
    \hfill
    \begin{subfigure}{0.5\textwidth}
        \centering
        \includegraphics[width=90mm,scale=0.5]{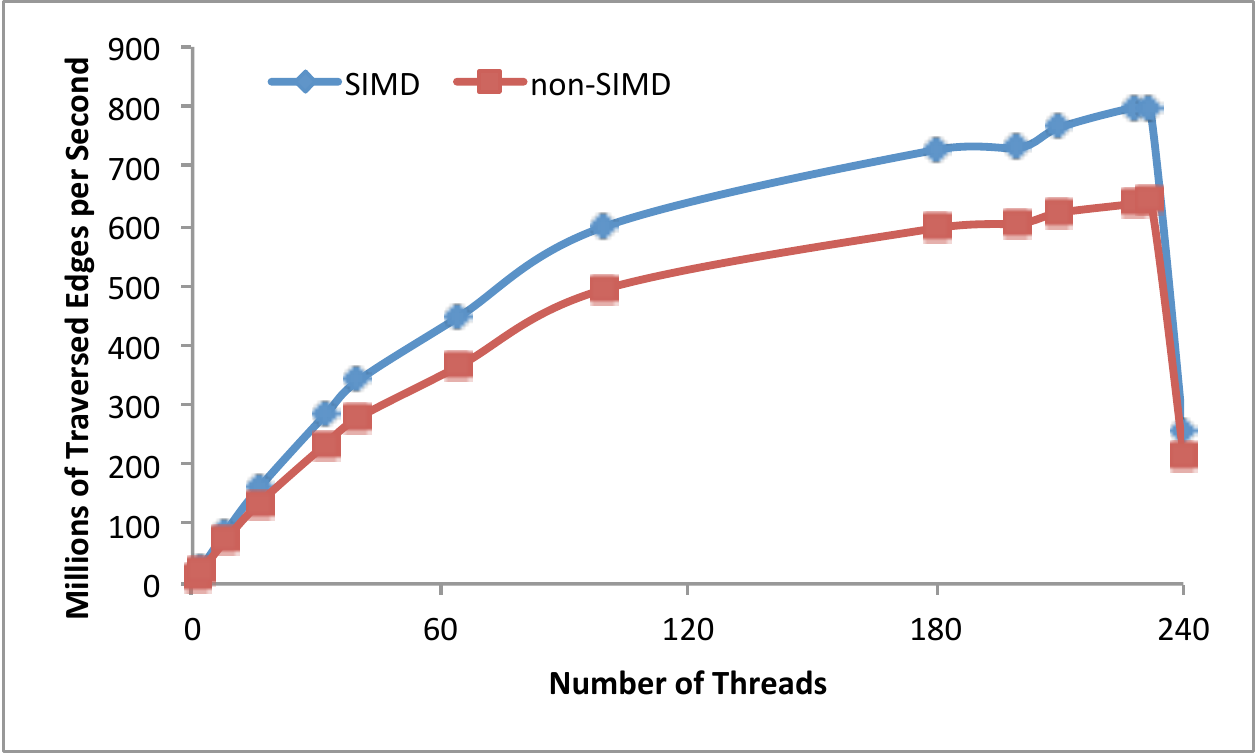}

                \caption{$SCALE=19$ and $edgefactor=16$.}
        \label{fig:simd}
    \end{subfigure}
    
        \begin{subfigure}{0.5\textwidth}
        \centering
        \includegraphics[width=90mm, scale=1.0]{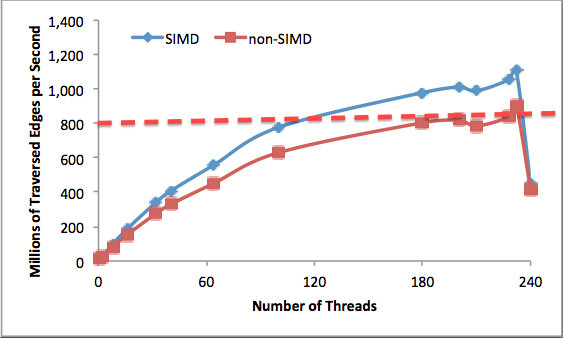}
        \caption{$SCALE=20$ and $edgefactor=16$.}
        \label{fig:simd}
    \end{subfigure}
     \caption{BFS \textit{non-SIMD} and SIMD experimental results for $SCALE$ values of 18, 19, 20 and $edgefactor=16$. In Figure (c) the dashed line represents the best MTEPS reported in \cite{MICgraphs}.}
    \label{fig:exp}
\end{figure}

%figura X -> mostrando differentes optimizations
%explicar el load imbalance
In Section \ref{sub:xeonphi} we present results illustrating the SIMD
optimizations. Figure \ref{fig:exp} shows the experimental results for
our \textit{non-simd} version and for the \textit{simd} optimized
version, which are described in Algorithm \ref{alg:parallelbfs}, for
three different graphs sizes, SCALE (18, 19 and 20), and edgefactor
16. Both versions present similar scalability but the simd version is
around 200 MTEPS faster than the non-simd one. However, as the number
of threads increases the rate of increase in TEPS decreases. This is a
result of hyprethreading and is discussed in Section
\ref{sub:affinity}. Finally, the variation in performance between 200
and 236 threads is due to workload imbalance during the exploration of
the vertices in each layer since, as the number of threads increases,
the chances of vertices processed by a thread having an uneven number
of adjacent vertices increases. The results show that our maximum
number of TEPS is above 1 gigatep. This is higher than that publishedf
in \cite{MICgraphs}, which gives approximately 800 MTEPS, for their
native BFS algorithm for the same graph size - the highest top-down
performance figure on a similar Xeon Phi that we have found in the
literature.

\subsection{Thread affinity}
\label{sub:affinity}
%We varied the number of threads pinned per core between 1, 2, 3 and 4 for the three different layout affinities: balanced, compact and scatter. %By experiment we selected balanced affinity since it was the one that shows better results. Balanced affinity allows to allocate threads are over %the cores in a way that threads pinned to the same core.
%Similarly, we ran executions for different number of cores from 36, 48. Both results were consistent. Table \ref{tab:affinity} shows the results, %varying the number of threads per core, for 48 cores for a graph size of SCALE=20 and edgefactor=16. 
Table \ref{tab:affinity} shows the result of running with 48 threads but varying the number of threads pinned per physical core manually.  
\begin{table}
\begin{center}
\caption{Performance \textit{SIMD} version by setting thread affinity for a graph size of SCALE=20 and edgefactor=16.}
\label{tab:affinity}
 \begin{tabular}{||c c c c||} 
 \hline
 \textbf{\#Threads} & \textbf{Thread Affinity} & \textbf{Cores} & \textbf{TEPS} \\ [0.5ex] 
 \hline
% 1 & 6 & 87837 & 787 \\ 
48&	1T/C & 48	& \textbf{4.69E+08} \\
 \hline
% 2 & 7 & 78 & 5415 \\
 	&2T/C& 24 & 2.67E+08 \\
% 3 & 545 & 778 & 7507 \\
 \hline
 &3T/C&16&	1.89E+08 \\
 \hline
 &4T/C&12&	1.42E+08 \\
 \hline
 \hline
\end{tabular}
\end{center}
\vspace{-8mm}
\end{table}
These results show the detrimental effects of over-populating the
cores for the BFS. The TEPS obtained with one and two threads per core
are significantly higher than the TEPS with three and four threads per
core. This reduction in the TEPS rate as the number of threads per
core increase is the key driver to the changes in slope observed in
Figure~\ref{fig:simd} occuring around 60, 120 and 180 cores when the
number of threads per core has to increase as more threads are
used. At these points, each thread's exclusive access to cache space
decreases as does its share of memory bandwidth.Despite these changes in slope, the performance of the both the simd
and non-simd BFS top-down algorithm continues to scale. So, by using
fully populated cores (59), each one with the maximum number of
threads, the number of TEPS for 236 logical threads is the
fastest. Beyond 236 thread, threads are placed on the final core,
which is reserved for the operating system on the Phi, resulting in a
dramatic fall in performance.

In the future, it might be possible to exploit this behaviour by
under-populating cores with threads performing BFS and to make use of
so-called {\em helper} threads running on a core to assist with, for
example, prefetching to help hide memory latency
\cite{Kamruzzaman:2011}.

\section{Related Work}
\label{relwork}
%bitmaps
This section briefly presents related work targeting mainly the
top-down BFS on the Xeon Phi. The Intel MIC Xeon Phi is a relative
recent parallel architecture released in 2012. An early work related
with graph algorithms on the Xeon Phi was reported
\cite{SauleC12}. However, they port a BFS parallel algorithm without
actually using the vector unit of the Xeon Phi. Other, and most
recent, work which is more related with ours is \cite{Gao2013} and
\cite{MICgraphs}. In \cite{Gao2013} they present the vectorization of
the top-down BFS algorithm, whereas in \cite{MICgraphs}, a complete
hybrid (top-down and bottom-up) heterogeneous BFS algorithm is
presented. However, few details about their use of vectorization are
presented. \cite{stanic2014} also explored the top-down BFS on the Xeon Phi but with a traditional, queue-based, algorithm that uses atomic updates. Despite their exploration related with the use of the vector unit and prefetching, their results are much lower than ours. We implemented our BFS top-down algorithm using SIMD
instructions by extending their work, which lead us to obtain a faster
vectorized implementation. Another approach to exploiting the Xeon
Phi is presented in a recently published system for heterogeneous
graph processing \cite{ChenHRJA15} allows the utilization of the Xeon
Phi in combination with a multi-core CPU through the use of a simple
programming interface. Although it is a good start towards
heterogeneous systems working on graphs, it is not made clear in this
paper how fast the resulting implementation is in comparison with
previous BFS implementations on the Xeon Phi.

%Different approaches to accelerate the BFS have been explored such as
%the use of GPUs and FGPAs but these are beyond the scope of this paper.

%\cite{SIMDTech} is a very useful guide to use the vector unit as it contains key recommendations for techniques to obtain good performance. These include: alignment, handling non-full vectorization and prefetching. In addition \cite{JhaHLCH15} presented a study of hash joins on the Xeon Phi. Their description of how to use the vector unit give us useful hints.

%meter paper - capitulo Auto-Vectorization Techniques for Modern SIMD Architectures Dresden

%\section{Conclusions and Future work}
\section{Conclusions}
\label{sec:conc}
In this paper, we revisited the vectorization and performance issues
of the top-down BFS algorithm on the Xeon Phi. In particular, we
studied the BFS top-down algorithm without (bit-wise) race conditions
to improve vectorization. The contributions of the paper are, first,
the development of an improved OpenMP parallel, highly vectorized SIMD
version of the BFS using vector intrinsics and successfully exploiting
data alignment and prefetching. This new implementation achieves a
higher number of TEPS than previous published results for the same
type of Xeon Phi. The second contribution is an investigation into the
impact of the thread affinity mapping and hyperthreading on
performance on an underpopulated system, a topic under researched in
the literature. This work suggests future possibilities to take
advantage of under-populating the cores and using the spare capacity
to improve latency hiding through the use of helper threads, for
example, to complement prefetching.

In the future, we plan to explore the benefit of our vectorization techniques beyond their use in native mode on the Xeon Phi, including targeting offload mode, GPGPUs and in the context of SSE and AVX SIMD technologies on multicore CPU-based systems. In addition, we are working on a version of the state-of-the-art hybrid BFS algorithm.

\section*{Acknowledgment}
This research was conducted with support from the UK Engineering and Physical Sciences Research Council (EPSRC) PAMELA EP/K008730/1, and AnyScale Apps EP/L000725/1. M. Paredes is funded by a National Council for Science and
Technology of Mexico PhD Scholarship. Mikel Luj{\'a}n is supported by a Royal Society University Research Fellowship.  

 %The following two commands are all you need in the
\bibliographystyle{abbrv}

\bibliography{refs}

\end{document}